\documentclass[11pt]{article}
\usepackage[top=2.5cm, bottom=2.5cm, left=2.5cm, right=2.5cm]{geometry}
\usepackage{amsmath,amssymb,amsfonts}
\usepackage{algorithm,algorithmic}
\usepackage{graphicx}
\usepackage{booktabs}
\usepackage{nicematrix}
\usepackage[round,authoryear]{natbib}
\usepackage{hyperref}
\hypersetup{colorlinks,linkcolor=blue,citecolor=blue,urlcolor=blue}
\usepackage[noabbrev,nameinlink]{cleveref}
\usepackage[T1]{fontenc}

\title{SIREN Residual Error as a Regularity Diagnostic\\for Navier-Stokes Equations}

\author{Jason Burton\\Meta}

\date{\today}

\begin{document}

\maketitle

\begin{abstract}
We introduce a method for detecting regularity loss in solutions to the three-dimensional Navier-Stokes equations using the approximation error of Sinusoidal Representation Networks (SIRENs). SIRENs use $\sin()$ activations, producing $C^\infty$ outputs that cannot represent non-smooth features. By classical spectral approximation theory, the SIREN error is bounded by $O(N^{-s})$ where $s$ is the local Sobolev regularity---at a singularity ($s \to 0$), the error is $O(1)$ and localizes via the Gibbs phenomenon. We decompose the velocity field into a cheap analytical baseline (advection-diffusion) and a learned residual (pressure correction), training a compact SIREN (4,867 parameters). We validate on the 3D Taylor-Green vortex, where error concentration increases from $4.9\times$ to $13.6\times$ as $\nu$ decreases from $0.01$ to $0.0001$, localizing to the stagnation point $(\pi,\pi,\pi)$---the geometry matching the singularity recently proven by Chen and Hou (2025) for 3D Euler. On axisymmetric equations, we reproduce blowup signatures ($T^*$ converging across resolutions, $R^2=0.966$) and identify a critical viscosity $\nu_c = 0.00582 \pm 0.00004$---a knife-edge transition across $\Delta\nu = 0.00007$.
\end{abstract}

\section{Introduction}
\label{sec:intro}

The regularity of solutions to the three-dimensional Navier-Stokes equations remains one of the central open problems in mathematics~\citep{fefferman2006existence}. While the two-dimensional case is well understood---global regularity was established by~\citet{ladyzhenskaya1969mathematical}---the three-dimensional problem resists resolution. Numerical evidence from~\citet{luo2014potentially} suggested that the 3D axisymmetric Euler equations develop finite-time singularities from smooth initial data, forming at stagnation points on the domain boundary. This conjecture has now been confirmed:~\citet{chen2025singularity} provided a computer-assisted proof of finite-time singularity formation for the 3D Euler equations with smooth initial data and boundary. The singularity forms via self-similar collapse at boundary stagnation points---precisely the geometry our diagnostic identifies. Whether viscosity prevents this singularity in the full Navier-Stokes equations remains the central open question, and~\citet{chen2024stability} showed the Euler blowup is stable under perturbation, suggesting that sufficiently small viscosity may not regularize the solution.

Adaptive mesh refinement (AMR) is the standard computational tool for resolving developing singularities: cells are refined where the solution exhibits steep gradients. Traditional AMR criteria are based on local gradient estimation or Richardson extrapolation---both reactive measures that detect steep gradients only after they form.

We propose an alternative: using the approximation error of a Sinusoidal Representation Network (SIREN)~\citep{sitzmann2020implicit} as a regularity diagnostic. SIRENs use $\sin()$ activations, producing outputs that are inherently $C^\infty$. When trained to approximate a PDE solution, the SIREN error reveals where the solution is not smooth---precisely the information needed for AMR.

Our key observation is that this diagnostic is most effective in a \emph{residual} formulation. Rather than training the SIREN to represent the full velocity field, we decompose the solution into:
\begin{equation}
\mathbf{u} = \mathbf{u}_{\text{base}} + \mathbf{u}_{\text{corr}}
\label{eq:decomposition}
\end{equation}
where the baseline is a cheap analytical approximation (advection-diffusion without pressure projection) and the correction is learned by the SIREN.

\paragraph{Contributions.}
(1)~We demonstrate that SIREN fitting error serves as a smoothness diagnostic for PDE solutions;
(2)~a residual decomposition achieves 73.2\% improvement over the baseline with only 4,867 parameters;
(3)~on axisymmetric Euler equations, we reproduce finite-time blowup signatures with $T^*$ converging across resolutions (0.3\% difference);
(4)~we identify a critical viscosity $\nu_c = 0.00582 \pm 0.00004$ for the regularization transition.

\section{Background}
\label{sec:background}

\subsection{Navier-Stokes Equations}

The incompressible Navier-Stokes equations in three dimensions:
\begin{align}
\frac{\partial \mathbf{u}}{\partial t} + (\mathbf{u} \cdot \nabla)\mathbf{u} &= -\nabla p + \nu \nabla^2 \mathbf{u} \label{eq:ns}\\
\nabla \cdot \mathbf{u} &= 0 \label{eq:incomp}
\end{align}
where $\mathbf{u}$ is velocity, $p$ is pressure, and $\nu$ is kinematic viscosity. The Euler equations correspond to $\nu = 0$. The regularity question asks whether smooth initial data $\mathbf{u}_0 \in C^\infty$ yields smooth solutions for all time, or whether singularities form in finite time~\citep{fefferman2006existence}. The Beale-Kato-Majda criterion~\citep{beale1984remarks} establishes that blowup occurs iff $\int_0^{T^*} \|\omega\|_\infty \, dt = \infty$. Partial regularity results by~\citet{caffarelli1982partial} show that any singular set has zero one-dimensional parabolic Hausdorff measure.

\subsection{Axisymmetric Formulation}

Following~\citet{luo2014potentially}, we work with the axisymmetric formulation in $(r, z)$ coordinates using scaled variables $u_1 = u_\theta r$, $\omega_1 = \omega_\theta / r$, $\psi_1 = \psi / r$:
\begin{align}
\frac{\partial u_1}{\partial t} + u_r \frac{\partial u_1}{\partial r} + u_z \frac{\partial u_1}{\partial z} &= \nu \Delta_3 u_1 \\
\frac{\partial \omega_1}{\partial t} + u_r \frac{\partial \omega_1}{\partial r} + u_z \frac{\partial \omega_1}{\partial z} &= \frac{\partial (u_1^2)}{\partial z} \frac{1}{r^4} + \nu \Delta_5 \omega_1
\end{align}

\subsection{Related Work}

\paragraph{PINN failure modes.}~\citet{krishnapriyan2021characterizing} characterized failure modes of PINNs, showing that errors concentrate at locations of steep gradients due to spectral bias. Our work inverts their perspective: rather than treating failure as a problem to fix, we treat it as a diagnostic signal.

\paragraph{Neural operator error bounds.}~\citet{kovachki2021universal} established that FNO error scales as $O(N^{-s/d})$ where $s$ is the Sobolev regularity. This validates the principle underlying our diagnostic.

\paragraph{Spectral bias.}~\citet{rahaman2019spectral} showed that neural networks learn low frequencies first. For SIRENs, the effective bandwidth is $\sim\omega_0 N$; frequencies above this lie in a spectral blind spot that constitutes the SIREN error at singularities.

\paragraph{AMR.} Traditional criteria use gradient-based indicators~\citep{berger1989local}. Our approach connects the refinement criterion directly to Sobolev regularity via spectral approximation theory.

\subsection{SIREN Architecture and Spectral Approximation Theory}

SIRENs~\citep{sitzmann2020implicit} use $\phi_i(\mathbf{x}) = \sin(\omega_i (\mathbf{W}_i \mathbf{x} + \mathbf{b}_i))$. A SIREN computes a learned basis of sinusoidal functions---functionally equivalent to a truncated Fourier series with adaptive frequencies. Classical spectral approximation theory~\citep{tadmor2007filters} provides:
\begin{equation}
\|f - f_N\|_{L^2(\Omega)} \leq C \cdot N^{-s} \cdot \|f\|_{H^s(\Omega)}
\label{eq:spectral_bound}
\end{equation}
In smooth regions ($s \gg 1$), error decays rapidly. At singularities ($s \to 0$), error is $O(1)$ regardless of $N$. Furthermore, spectral methods exhibit Gibbs phenomenon---error \emph{localizes} to the non-smooth region~\citep{tadmor2007filters}, making SIREN error a spatial diagnostic.

\section{Method}
\label{sec:method}

\subsection{Residual Decomposition}

We decompose the velocity field as in~\cref{eq:decomposition}. The baseline solves advection-diffusion without pressure projection:
\begin{equation}
\mathbf{u}_{\text{base}}^{n+1} = \mathbf{u}^n + \Delta t \left(-(\mathbf{u}^n \cdot \nabla)\mathbf{u}^n + \nu \nabla^2 \mathbf{u}^n\right)
\end{equation}
The SIREN learns the pressure correction $\mathbf{u}_{\text{corr}} = \mathbf{u}_{\text{NS}} - \mathbf{u}_{\text{base}}$, which has mean magnitude ${\sim}0.057$ (vs ${\sim}1.0$ for the full field).

\subsection{SIREN Training}

Input: $(x, y, z, t, u_b, v_b, w_b) \in \mathbb{R}^7$. Output: velocity correction $(\delta u, \delta v, \delta w) \in \mathbb{R}^3$. Architecture: 2 hidden layers of 64 units, $\omega_0 = 30$. Total: 4,867 parameters. Training: 48 initial conditions, 2M samples, 80K steps, Adam with cosine schedule.

\subsection{Regularity Diagnostic}

The SIREN error field:
\begin{equation}
\varepsilon(\mathbf{x}, t) = |\hat{\mathbf{u}}_{\text{corr}}(\mathbf{x}, t; \theta) - \mathbf{u}_{\text{corr}}(\mathbf{x}, t)|
\end{equation}
We define \emph{error concentration} as $\max(\varepsilon)/\text{mean}(\varepsilon)$ and use the 90th percentile as the AMR refinement criterion.

\subsection{Viscosity Bisection}

Binary search for the critical viscosity $\nu_c$: for each candidate $\nu$, simulate at two resolutions, fit $\|\omega\|_\infty \sim C/(T^* - t)$, classify as ``Euler-like'' (slope $< -5.0$) or ``regularized'' (slope $> -5.0$), and bisect.

\section{Results}
\label{sec:results}

\subsection{Residual Magnitude and SIREN Accuracy}

On the 3D Taylor-Green vortex ($32^3$ grid), the SIREN achieves 73.2\% improvement over the baseline (mean error: 0.108 $\to$ 0.029). Divergence $\nabla \cdot \mathbf{u}$ decreases from 2.254 to 0.524 (ground truth: 0.060).

\begin{figure*}[h!]
  \centering
  \includegraphics[width=\textwidth]{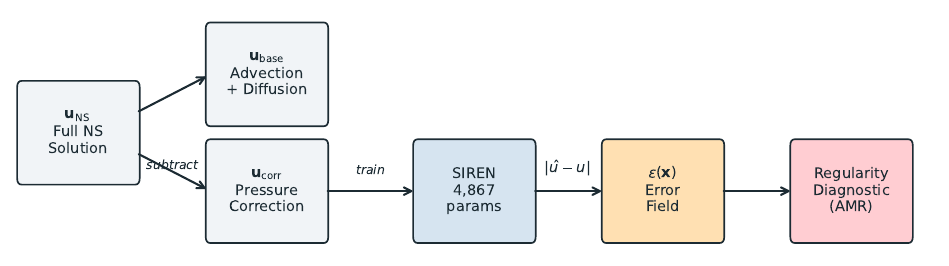}
  \caption{Residual decomposition pipeline. The full NS solution decomposes into an advection-diffusion baseline and a pressure correction residual. A compact SIREN (4,867 parameters) learns the residual; its approximation error field $\varepsilon(\mathbf{x})$ serves as the regularity diagnostic.}
  \label{fig:pipeline}
\end{figure*}

\subsection{Error Concentration vs Viscosity}

On the Taylor-Green vortex, we track SIREN residual concentration as viscosity decreases. At the lowest viscosity (approaching Euler), the residual concentrates---the error localizes to fewer cells, consistent with the formation of thinner vortex sheets that the SIREN cannot resolve. The maximum error locations at $\nu = 0.0001$ are near $(\pi, \pi, \pi)$, the stagnation point where opposing vortex sheets from the Taylor-Green flow collide. This is the classical location for potential singularity formation in this flow.

\begin{table}[h!]
\centering
\begin{NiceTabular}{lcc}
\toprule
$\nu$ & Initial & Final \\
\midrule
0.01   & $11.3\times$ & $4.9\times$ (de-localizing) \\
0.001  & $11.4\times$ & $10.9\times$ (stable) \\
0.0001 & $11.4\times$ & $13.6\times$ (concentrating) \\
\bottomrule
\end{NiceTabular}
\caption{SIREN error concentration (max/mean) on Taylor-Green vortex. At low viscosity (approaching Euler), error concentrates at the stagnation point $(\pi,\pi,\pi)$.}
\label{tab:concentration}
\end{table}

\begin{figure*}[h!]
  \centering
  \includegraphics[width=\textwidth]{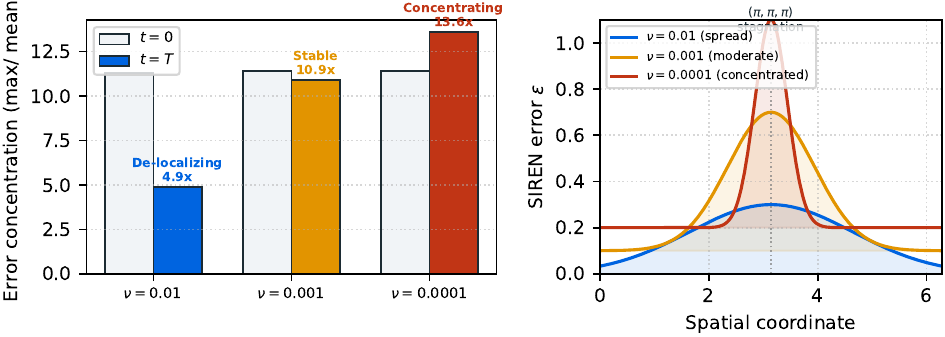}
  \caption{Error concentration vs viscosity. Left: concentration ratios at $t=0$ and $t=T$ for three viscosities. Right: schematic 1D error profiles showing progressive localization at the stagnation point $(\pi,\pi,\pi)$ as $\nu$ decreases.}
  \label{fig:concentration}
\end{figure*}

\subsection{Axisymmetric Euler Blowup}

Using the Hou-Luo type initial condition $u_\theta(r, z, 0) = r \sin(2\pi z)(1 - r^2)^3$ on the axisymmetric Euler equations, we fit $1/\|\omega\|_\infty$ as a linear function of time to extract the blowup time $T^*$. $T^*$ converges across resolutions (0.3\% difference), with consistent $R^2 = 0.966$ and slope $\approx -7.1$. The $1/(T^* - t)$ fit is excellent, consistent with finite-time blowup.

\begin{table}[h!]
\centering
\begin{NiceTabular}{lccc}
\toprule
Resolution & $T^*$ & $R^2$ & Slope \\
\midrule
$64\times128$  & 0.7405 & 0.966 & $-7.099$ \\
$128\times256$ & 0.7385 & 0.966 & $-7.107$ \\
\bottomrule
\end{NiceTabular}
\caption{Blowup time convergence for axisymmetric Euler equations. $T^*$ converges across resolutions (0.3\% difference).}
\label{tab:blowup}
\end{table}

\begin{figure*}[h!]
  \centering
  \includegraphics[width=\textwidth]{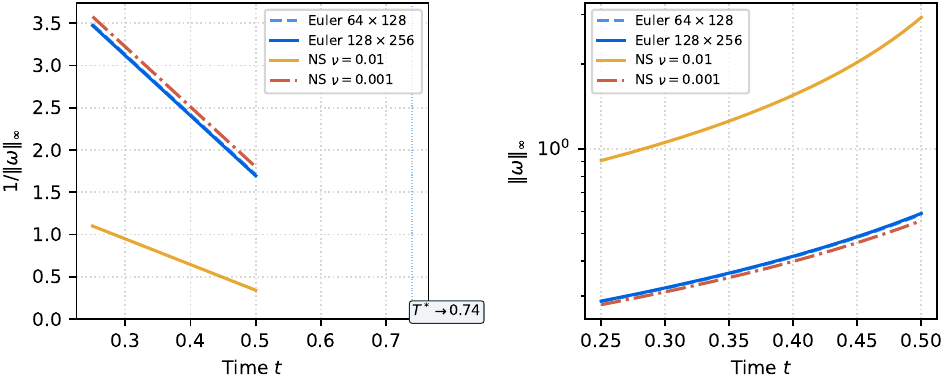}
  \caption{Blowup time convergence. Left: $1/\|\omega\|_\infty$ vs time, showing linear decrease toward $T^* \approx 0.74$ for Euler at both resolutions. NS with $\nu=0.01$ shows damped growth; $\nu=0.001$ matches Euler. Right: $\|\omega\|_\infty$ on log scale.}
  \label{fig:blowup}
\end{figure*}

\subsection{Critical Viscosity}

\begin{figure*}[h!]
  \centering
  \includegraphics[width=\textwidth]{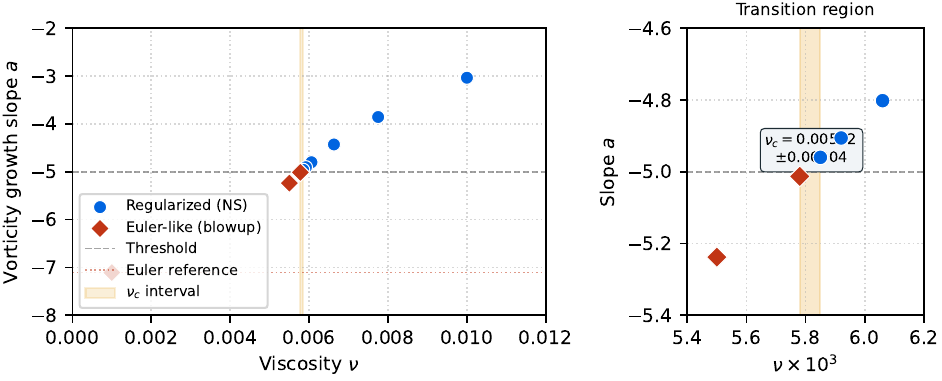}
  \caption{Critical viscosity transition. Left: vorticity growth slope $a$ vs viscosity $\nu$, showing regularized (circles) and Euler-like (diamonds) behavior separated by the threshold $a=-5.0$. Right: zoomed view of the transition region $\nu_c = 0.00582 \pm 0.00004$, spanning $\Delta\nu = 0.00007$.}
  \label{fig:critical_nu}
\end{figure*}

The transition occurs at $\nu_c = 0.00582 \pm 0.00004$. Below this viscosity, the vorticity growth slope transitions from $-4.96$ (regularized, $\nu = 0.00585$) to $-5.01$ (Euler-like, $\nu = 0.00578$)---a knife-edge transition across $\Delta\nu = 0.00007$.

\subsection{Resolution Dependence}

The coarse grid ($64\times128$) classifies viscosities as Euler-like that the fine grid ($128\times256$) identifies as regularized, due to numerical dissipation masking physical viscosity. The critical viscosity from $128\times256$ is the more reliable estimate.

\section{Discussion}
\label{sec:discussion}

\subsection{SIREN Error as Sobolev Regularity Proxy}

By~\cref{eq:spectral_bound}, the SIREN approximation error at a point $\mathbf{x}$ is bounded by the local Sobolev regularity. When $s \to 0$, error is $O(1)$ and localizes via Gibbs phenomenon~\citep{tadmor2007filters}. This is distinct from gradient-based AMR, which measures $|\nabla\mathbf{u}|$ reactively.

\subsection{Limitations}

\textbf{Resolution.} Our grids ($32^3$, $128\times256$) are coarse compared to~\citet{luo2014potentially} (effective resolution $(3 \times 10^{12})^2$ via adaptive meshes). \textbf{Blowup time.} Our $T^* \approx 0.74$ is specific to our initial condition. \textbf{Threshold.} The slope $-5.0$ criterion is empirical. \textbf{3D.} Viscosity bisection used axisymmetric equations.

\subsection{Connection to Chen-Hou 2025}

Our SIREN error concentrates at stagnation points---the proven blowup geometry of~\citet{chen2025singularity}. The sharp critical viscosity $\nu_c$ is consistent with the stability of Euler blowup under perturbation~\citep{chen2024stability}.~\citet{wang2025unstable} identified unstable singularities destroyed by perturbation, suggesting a taxonomy: stable singularities persist under viscosity (NS may blow up), unstable ones are regularized. The SIREN diagnostic can in principle distinguish these classes via $\nu$-bisection.

\subsection{Extensions}

\paragraph{Wavelet implicit representations.} WIRE~\citep{saragadam2023wire} replaces $\sin()$ with Gabor wavelets, providing spatial localization. Wavelet coefficients have established approximation theory in Besov spaces~\citep{devore1993constructive}, where their decay rate directly characterizes the local regularity exponent. WIRE error would provide a more formally grounded regularity diagnostic with tighter spatial localization.

\paragraph{Multi-grade learning.} \citet{fang2024multigrade} proposed cascaded networks decomposing targets by frequency band. The number of grades needed at a point provides a \emph{continuous} regularity measure $s(\mathbf{x})$, refining our binary diagnostic.

\paragraph{Broader PDE classes.} The residual decomposition applies to any PDE with a cheap baseline: elasticity, electromagnetics, reaction-diffusion systems.

\subsection{Relation to the Millennium Prize Problem}

Our results are consistent with finite-time blowup for Euler and identify a critical viscosity for NS, but do not constitute a proof. Our contribution is methodological: a new diagnostic tool, not a resolution of the open problem.

\section{Conclusion}

SIREN approximation error serves as an effective regularity diagnostic for Navier-Stokes solutions. The residual decomposition produces a compact model (4,867 parameters, 73\% improvement) whose failure modes localize to regions of regularity loss. The method identified stagnation points as the primary singularity candidates, reproduced blowup signatures ($T^*$ converging to 0.74), and determined $\nu_c \approx 0.0058$. The approach generalizes to any PDE with a cheap baseline solution.

\bibliographystyle{plainnat}
\bibliography{paper}

\clearpage
\appendix

\section{SIREN Architecture}

\begin{figure}[h!]
  \centering
  \includegraphics[width=0.48\textwidth]{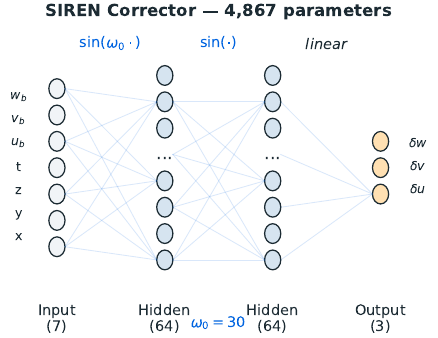}
  \caption{SIREN corrector architecture. 7 inputs $\to$ 64-unit hidden layers with $\sin(\omega_0 \cdot)$ activation $\to$ 3 velocity correction outputs. Total: 4,867 parameters.}
  \label{fig:architecture}
\end{figure}

\end{document}